\documentclass[aps,prd,amsfonts,amsmath,nofootinbib,tightenlines,preprint,showpacs,longbibliography]{revtex4-1}
\pdfoutput=1

\usepackage{graphicx}
    \graphicspath{ {./figures/} }
\usepackage{amsmath}
\usepackage{array}
\def\oA{A_0}
\def\oB{B_0}
\def\bA{\bar A}
\def\bB{\bar B}
\def\bs{\bar s}
\def\bu{\bar u}
\def\bv{\bar v}
\def\bX{\bar X}
\def\bg{\bar g}
\def\dX{\Delta X}
\def\dv{\Delta v}
\def\du{\Delta u}
\def\dA{\Delta A}
\def\dB{\Delta B}
\def\umax{u_{\text{max}}}
\def\lu{\lambda_0}

\def\oAB{\oA''\cdot\oB''}

\def\interval{\mathcal{I}}
\def\intervalv{\interval_{,v}}
\def\intervalvv{\interval_{,vv}}
\def\intervalu{\interval_{,u}}
\def\order{\mathcal{O}}
\def\sab{s_{\alpha\beta}}
\def\Sab{\Sigma_{\alpha\beta}}

\begin{document}

\title{Gravitational back-reaction near cosmic string kinks and cusps}
\author{Jose J. Blanco-Pillado}
\email{josejuan.blanco@ehu.es}
\affiliation{Department of Theoretical Physics, UPV/EHU,\\48080, Bilbao, Spain}
\affiliation{IKERBASQUE, Basque Foundation for Science, 48011, Bilbao, Spain}
\author{Ken D. Olum}
\email{kdo@cosmos.phy.tufts.edu}
\affiliation{Institute of Cosmology, Department of Physics and Astronomy,\\Tufts University, Medford, MA 02155, USA}
\author{Jeremy M. Wachter}
\email{jeremy.wachter@ehu.es}
\affiliation{Department of Theoretical Physics, UPV/EHU,\\48080, Bilbao, Spain}

\begin{abstract}

We find the leading-order effect of gravitational back-reaction on
cosmic strings for points near kinks and cusps.  Near a kink, the
effect diverges as the inverse cube root of the distance to the kink,
and acts in a direction transverse to the worldsheet.  Over time the
kink is rounded off, but only regions fairly close to the kink are
significantly affected.  Near cusps, the effect diverges inverse
linearly with the distance to the cusp, and acts against the direction
of the cusp motion. This results in a fractional loss of string energy
that diverges logarithmically with the distance of closest approach to
the cusp.

\end{abstract}

\maketitle

\section{Introduction}\label{sec:intro}

Cosmic strings are one-dimensional topological defects which may form
dynamically at a symmetry breaking phase transition in the early
universe~\cite{Kibble:1976sj,Vilenkin:2000jqa}. Models of string
theory also suggest the possibility that fundamental strings (and
D1-branes) can be stretched by the cosmic expansion in the early
universe and form a cosmic superstring network
\cite{Dvali:2003zj,Copeland:2003bj}. As massive objects generically in
motion, the strings radiate gravitational waves, and a network of
cosmic string loops would produce a stochastic background (E.g.,
see~\cite{Blanco-Pillado:2017oxo} and references therein).  They are
therefore of great interest to gravitational wave observatories, many
of which are actively searching for cosmic
strings~\cite{Arzoumanian:2018saf,Abbott:2017mem,Lentati:2015qwp}.

The emission of gravitational waves is accompanied by back-reaction:
cosmic strings self-interact gravitationally, which generically
changes their shape and has the potential to affect the stochastic
gravitational wave background. However, owing to the complexity of a
typical cosmic string loop's shape~\cite{Blanco-Pillado:2015ana}, it
is generally infeasible to solve analytically for the evolution of a
cosmic string undergoing gravitational back-reaction. Analytic
solutions are known only for a few simple loop
shapes~\cite{0264-9381-22-13-002,Wachter:2016rwc}.

Instead we focus here on the self-interaction process very near
features of the cosmic string loop of particular interest to its
overall evolution: kinks and cusps. Kinks are persistent points on a
loop where there is a discontinuity in the the tangent vector to the
loop~\cite{Garfinkle:1987yw}; cusps are transient points that recur
once per oscillation period where the string moves (formally) at
the speed of light~\cite{Turok:1984cn}.

The pioneering work in cosmic string back-reaction was done by
Quashnock and Spergel~\cite{Quashnock:1990wv}.  They found that there
were no divergences in the gravitational back-reaction due to nearby
points on a smooth string.  However, in the case of kinks and cusps,
the string is not smooth, so their argument does not apply, and
there is the possibility of effects that become unboundedly large at
points arbitrarily close to these features.

Indeed, we find that points on cosmic strings very near to kinks and
cusps experience a divergent self-force.  This corrects the claim made
by two of us (J.M.W and K.D.O) in Ref.~\cite{Wachter:2016hgi} that the
back-reaction near kinks was not divergent and thus that kinks would
not be rounded off.  The error in the analysis of
Ref.~\cite{Wachter:2016hgi} is discussed in its erratum, found in its
arXiv ancillary files.

In Sec.~\ref{sec:setup}, we frame the problem and establish our
methodologies. In Sec.~\ref{sec:generic}, we find the self-interaction
for a generic point far from kinks or cusps, reproducing a result of
Ref.~\cite{Quashnock:1990wv}.  In Sec.~\ref{sec:kink}, we solve for
the self-interaction very near to a kink, and in Sec.~\ref{sec:cusp}
for very near to a cusp. We conclude in Sec.~\ref{sec:conclusions}.

We work in linearized gravity, which is accurate because the string's
coupling to gravity is very small. Our metric signature is $(-+++)$,
and we work in units where the speed of light is one.

\section{Setup}\label{sec:setup}
\subsection{The string worldsheet}\label{sec:worldsheet}
We first consider a string following the Nambu-Goto equations of motion in
flat space.  As usual, we will describe the string in the conformal
gauge and choose the timelike parameter on the string equal to the
spacetime coordinate $t$.  Then the string motion is given by~\cite{Vilenkin:2000jqa},
\begin{equation}
    X^\gamma = \frac{A^\gamma(v)+B^\gamma(u)}{2}\,,
\end{equation}
where $u$ and $v$ are null coordinates and $A' = dA/dv$ and $B' =
dB/dv$ are null vectors tangent to the string worldsheet and with unit
time component.  In terms of the usual spacelike string coordinate
$\sigma$ that parameterizes energy, $u=t+\sigma$ and $v=t-\sigma$.

The gravitational effect of the string will give rise to a small
perturbation to the metric, which will in turn give a small correction
to the string motion.  We will compute that correction and apply it
after a complete oscillation by changing the functions $A$ and $B$.
We will see below that this approximation is very accurate in
realistic situations.

The tangent vectors $A'$ and $B'$ have unit spatial length, and
so we commonly represent their spatial parts, $\mathbf{A}'(v)$ and
$\mathbf{B}'(u)$, as curves on the unit sphere
\cite{Kibble:1982cb}.  In this
representation, we may easily identify kinks and cusps: kinks are
discontinuous jumps of a tangent vector from one point on the unit
sphere to another, while cusps are points on the unit sphere where the
tangent vector curves cross. Kinks are a phenomenon due to one of the
two tangent vectors, and are present at any time slice of the loop,
while cusps involve both tangent vectors and only appear at a specific
moment in each oscillation. This representation of kinks and cusps
demonstrates that kinks inhibit cusps: a discontinuous jump in a
tangent vector's curve allows it to avoid an intersection with the
other tangent vector. For a closed loop in the rest frame, the
``center of mass'' of the tangent vector curves must lie at the center
of the unit sphere, and so string loops will generically have cusps
unless they contain kinks.

We are interested in the back-reaction on some point on a
cosmic string, which we will refer to as the \emph{observation point}
or simply the \emph{observer}. We will indicate observer quantities by
an overbar, i.e. the observer is located at $\bX$. 

In most cases we place the origin of coordinates at the observer,
but for observers near a cusp, we will use the cusp itself as the
origin.  Quantities at the origin will be denoted by subscript $0$,
and we will expand around that point,
\begin{subequations}\label{eqn:taylor}\begin{align} 
    A(v) = vA'_0 + \frac{v^2}{2}A''_0 + \frac{v^3}{6}A'''_0\,,\\
    B(u) = uB'_0 + \frac{u^2}{2}B''_0 + \frac{u^3}{6}B'''_0\,.
\end{align}\end{subequations}
In order for the vectors to be null ($A'\cdot A' = B'\cdot
B'=0$), we must introduce the constraints
\begin{subequations}\label{eqn:null-constraints}\begin{align}
   \oA'\cdot\oA'&=0\,,\\	
    \oA'\cdot\oA''&=0\,,\\
    \oA'\cdot\oA'''&=-\oA''^2\,,
\end{align} \end{subequations}
and likewise in $B$.

The acceleration felt by a point due to the gravitational
effect of the string is, at first order,~\cite{Quashnock:1990wv}
\begin{equation}\label{eqn:xuv}
    \bX^\gamma_{,uv} = -\frac18\eta^{\gamma\rho}\left(h_{\beta\rho,\alpha}+h_{\rho\alpha,\beta}-h_{\alpha\beta,\rho}\right)\bA'^\alpha\bB'^\beta\,.
\end{equation}
Here $\eta_{\mu \nu}$ is the flat-space metric, and $h_{\alpha
  \beta}$ is the perturbation to that metric.  We can compute the
change of the tangent vectors due to gravitational back-reaction by
integrating the acceleration induced by the unperturbed
worldsheet\footnote{This is the approximation that was used
  in~\cite{Quashnock:1990wv,Wachter:2016hgi,Wachter:2016rwc}.} over a
full oscillation,
\begin{subequations}\label{eqn:delta-nv} \begin{align}
    \dA'^\gamma &= 2\int_0^L X^\gamma_{,uv}du\,,\\
    \dB'^\gamma &= 2\int_0^L X^\gamma_{,uv}dv\,.
\end{align} \end{subequations}

The metric depends on the choice of coordinates (i.e., the gauge) for
the perturbed spacetime.  Thus $\bX^\gamma_{,uv}$ may contain gauge
artifacts.  However, $\dA'$ and $\dB'$ do not have this problem.  The
metric oscillates with the oscillation of the string, but $\dA'$ and
$\dB'$ grow linearly with the number of oscillations (as long as we
continue to use the approximation that the source worldsheet is
unchanged).  This provides a clean separation between effects that may
and those that may not have gauge dependence.

Since the corrections to $A'$ leave $A'$ null, we will
automatically have $\dA'\cdot A'= 0$.  But because of the Lorentzian
metric, adding $\dA'$ may change the length of $A'$, which represents
a loss of energy from the string.  Since we demand that
$|\mathbf{A}'|= 1$, we must change the parameterization by redefining
$v$.  The same remarks apply to $B'(u)$.

This reparameterization raises the question of whether the
divergences that we find below could be only artifacts of the
parameterization.  The answer is that so long as the divergent effect
changes the direction of $A'$ or $B'$, it is not a parameterization
artifact, because these changes of direction cannot be removed by
reparameterization.

\subsection{The metric perturbation}\label{ssec:h}

We will now compute the metric perturbation at an observer position
$\bX$ due to some source point $X$.  Let $\dX = X - \bX$, the vector from the observer
to the source, and let $\interval=(\dX)^2$, the squared interval
between source and observer.

Starting from the linearized Einstein equations,
\begin{equation}
    \Box h_{\alpha\beta} = 16\pi GS_{\alpha\beta}\,,
\end{equation}
where $G$ is Newton's constant and $S$ the trace-reversed
stress-energy tensor, we solve by the method of Green's functions,
\begin{equation}\label{eqn:hGreens}
    h_{\alpha\beta}(\bX)= 8G\int d^4x S_{\alpha\beta}(x)\delta(\interval)\,,
\end{equation}
where we  take the integral only
over source points $X$ in the past of $\bX$.  A string has a stress tensor of the form~\cite{Quashnock:1990wv}
\begin{equation}\label{eqn:Ss}
    S_{\alpha\beta}(X)=\frac{\mu}{4}\int du\,dv ~s_{\alpha\beta}~\delta^{(4)}(X-X(u,v))\,,
\end{equation}
where $\mu$ denotes the energy per unit length of the string and we have defined\footnote{The quantity $s_{\alpha\beta}$ here is twice the
  $\sigma_{\alpha\beta}$ of Ref.~\cite{Wachter:2016rwc} and four times
  the $F_{\alpha\beta}$ of Ref.~\cite{Quashnock:1990wv}.}
$s_{\alpha\beta}=\Sigma_{\alpha\beta}(A',B')$, with
\begin{equation}\label{eqn:Sigma}
    \Sigma_{\alpha\beta}(P,Q)=P_\alpha Q_\beta + Q_\alpha P_\beta - \eta_{\alpha\beta}(P\cdot Q)\,.
\end{equation}

We pause here to note two important features of $\Sigma$. If $N$ is a null vector,
\begin{subequations}\label{eqn:Sigma-features} \begin{align}
    \Sigma_{\alpha\beta}(N,Q)N^\alpha &= 0\,,\\
    \Sigma_{\alpha\beta}(P,Q)N^\alpha N^\beta &= 2(N\cdot P)(N\cdot Q)\,.
\end{align} \end{subequations}
These features will lead to a number of useful simplifications further down the road.

Putting Eq.~(\ref{eqn:Ss}) into Eq.~(\ref{eqn:hGreens}), we find
\begin{equation}\label{eqn:hab}
    h_{\alpha\beta}(\bX)= 2G\mu\int du\,dv ~s_{\alpha\beta}(X)~\delta(\interval)\,.
\end{equation}
The metric is thus determined by the effect of all places where the
backward lightcone from the observation point intersects the string
worldsheet, which we will call the \emph{intersection line}.

We can eliminate one integral in Eq.~(\ref{eqn:hab}) by changing
variables in the $\delta$-function.  For example, to eliminate $v$, we
write
\begin{equation}\label{eqn:deltachange}
  \delta(\interval)=-\frac{\delta(v-v(u))}{\intervalv}\,,
\end{equation}
where $v(u)$ denotes the (unique) value of $v$ for the given $u$ that
puts the point $(u, v)$ on the past lightcone of the observer.  (The
negative sign in Eq.~(\ref{eqn:deltachange}) appears because
$\intervalv < 0$).  The result will be an integral giving the metric
at $\bX$ as a sum of the contributions due to the stress-energy at
each source point.  We could then differentiate $h_{\alpha\beta}$ and
use Eq.~(\ref{eqn:xuv}) to find the acceleration.  Indeed, this is the
procedure used in Ref.~\cite{Wachter:2016rwc}.

In order to differentiate, though, we would need the metric not just on the
worldsheet but nearby.  It turns out to be easier to differentiate
Eq.~(\ref{eqn:hab}) first~\cite{Quashnock:1990wv},
\begin{equation}
    h_{\alpha\beta,\gamma}(\bX)= 4G\mu\int du\,dv ~s_{\alpha\beta}(X)~\delta'(\interval)X^\gamma\,.
\end{equation}
Then we can write the derivative with respect to $\interval$ in terms
of a derivative with respect to $v$ (say),
\begin{equation}
    h_{\alpha\beta,\gamma}(\bX)= 4G\mu\int du\,dv
  \left(  \frac{s_{\alpha\beta}(X)}{\intervalv}\right)\frac{\partial}{\partial v}
\delta(\interval),.
\end{equation}
We integrate by parts and then proceed as above to get~\cite{Quashnock:1990wv}
\begin{equation}\label{eqn:habg}
    h_{\alpha\beta,\gamma} (\bX)=4G\mu\int du \left[\frac{1}{\intervalv}\,\frac{\partial}{\partial v}\left(\frac{s_{\alpha\beta}\dX_\gamma}{\intervalv}\right)\right]_{v=v(u)}\,.
\end{equation}
Equation~(\ref{eqn:habg}) gives the metric derivative at $\bX$ as an
integral over source points and allows us to consider $\bX$ only on
the worldsheet.  We could also have chosen to convert $\delta'$ using
$u$ instead of $v$, and (independently) to change variables in
$\delta(\interval)$ to $u$ instead of $v$.

To apply Eq.~(\ref{eqn:habg}), we proceed as follows.  There are two
branches to the intersection line near $\bX$, one going mostly in the
direction of decreasing $u$ and the other mostly in the direction of
decreasing $v$.  We will consider only the former, meaning source
points where $\du = u-\bu < 0$ and $\dv = v-\bv \ge 0$.  The latter
condition is necessary because if $\du,\dv<0$, the source point would be
in chronological past of the observer, not on the lightcone.

Given the specific form of string, we can write an explicit expression for
$\interval(u, v)$.  For a specific $u < 0$, we can solve
$\interval=0$ for $v$.  We then perform the operations in
Eq.~(\ref{eqn:habg}) to find $h_{\alpha\beta,\gamma}$.

We can write
\begin{equation}
\interval = \left(\frac{\dA(v)+\dB(u)}{2}\right)^2
\end{equation}
so we have the derivative
\begin{equation}\label{eqn:Iv1}
\intervalv = \left(\frac{\dA(v)+\dB(u)}{2}\right)\cdot \dA'
\end{equation}

\subsection{Coordinate system}\label{ssec:uvcd}

We can simplify our calculations by using a coordinate system adapted
to the worldsheet.  For most purposes, we will use a pseudo-orthogonal
coordinate system $(u, v, c, d)$ constructed around the observation
point, with basis vectors $e_{(u)} =\bB'/2$, $e_{(v)} =\bA'/2$, and
$e_{(c)}$ and $e_{(d)}$ any unit spacelike vectors perpendicular to
$\bA'$ and $\bB'$ and to each other.  Defining $Z = \bar A'\cdot \bar B'$,
the corresponding covector basis
is $e^{(u)} =2\bA'/Z$, $e^{(v)} =2\bB'/Z$, $e^{(c)}= e_{(c)}$,
$e^{(d)}= e_{(d)}$, and the metric tensor is in $uvcd$ coordinates is
\begin{equation}\label{eqn:eta-uvcd}
    \eta_{\alpha\beta}  = \left(\begin{array}{cccc} 0 & Z/4 & 0 & 0\\ Z/4 & 0 & 0 & 0\\ 0 & 0 & 1 & 0\\ 0 & 0 & 0 & 1\end{array}\right)\,,\qquad\eta^{\alpha\beta}  = \left(\begin{array}{cccc} 0 & 4/Z & 0 & 0\\ 4/Z & 0 & 0 & 0\\ 0 & 0 & 1 & 0\\ 0 & 0 & 0 & 1\end{array}\right)\,.
\end{equation}

This basis allows us a number of simplifications in vector components. Namely,
\begin{itemize}
    \item $\bar A'^v=2$, $\bar B'^u=2$, and all other components of both are zero.
    \item $\bar A'_u=Z/2$, $\bar B'_v=Z/2$, and all other components of both are zero.
    \item Because $\bar A'\cdot \bar A''=0$, we have $\bar A''^u=\bar A''_v=0$, and similarly, $\bar B''^v=\bar B''_u=0$.
\end{itemize}
There is a cancellation in $\Sigma_{uv}$, so that
\begin{equation}
    \Sigma_{uv}(P,Q) = -\frac{Z}{4}(P\star Q)\,,
\end{equation}
where we define $P\star Q = P_cQ_c+P_dQ_d$, which can be understood as
the inner product in the subspace perpendicular to the worldsheet.

Finally, Eq.~(\ref{eqn:xuv}) becomes
\begin{equation}\label{eqn:xuv-poc}
    \bX^\gamma_{,uv} = -\frac12\eta^{\gamma\rho}\left(h_{u\rho,v}+h_{\rho v,u}-h_{uv,\rho}\right)\,.
\end{equation}
Then, making use of Eq.~(\ref{eqn:eta-uvcd}), we find the acceleration components in the $uvcd$ basis,
\begin{subequations} \label{eqn:x} \begin{align}
\label{eqn:xu} X^u_{,uv} &= -\frac{2}{Z}h_{vv,u}\,\\
\label{eqn:xv} X^v_{,uv} &= -\frac{2}{Z}h_{uu,v}\,\\
\label{eqn:xc} X^c_{,uv} &= \frac12(h_{uv,c}-h_{uc,v}-h_{vc,u})\,\\
\label{eqn:xd} X^d_{,uv} &= \frac12(h_{uv,d}-h_{ud,v}-h_{vd,u})\, ~.
\end{align} \end{subequations}

\section{Near a generic point}\label{sec:generic}

We will now find the leading-order effect of back-reaction on the
smooth string worldsheet, reproducing a result of Quashnock and
Spergel~\cite{Quashnock:1990wv}.  We will choose the origin at the
observation point.  Then
\begin{equation}
    \interval = \left(\frac{A(v)+B(u)}{2}\right)^2\,.
\end{equation}
We will consider the branch of the intersection line going nearly in
the $-u$ direction, so $|u|\gg|v|$.  To find $v(u)$, we use
Eqs.~(\ref{eqn:taylor},\ref{eqn:null-constraints}) and disregard terms
of order $v^2$, $u^2v$, $u^5$ and higher, to find
\begin{equation}\label{eqn:I1}
    \interval = \frac{Zuv}{2}-\frac{\bB''^2u^4}{48}\,.
\end{equation}
Setting $\interval = 0$ gives
\begin{equation}\label{eqn:vu-generic}
    v(u) = \frac{\bB''^2u^3}{24Z}\,.
\end{equation}
Thus we have consistently disregarded terms higher than order $u^4$ in
Eq.~(\ref{eqn:I1}).

We will need to be more accurate in computing $\intervalv$.  From
Eq.~(\ref{eqn:Iv1}), $\intervalv=A\cdot A'/2+B\cdot A'/2$.  But from
Eqs.~(\ref{eqn:taylor},\ref{eqn:null-constraints}) the first term will
be $\order(v^3)$. We will not be interested in effects at this level,
and so we can write
\begin{equation}\label{eqn:Iv2}
    \intervalv = \frac{B\cdot A'}{2} = uA'_u + \frac{(\bB''\cdot A')u^2}{4}\,.
\end{equation}
Higher orders in $u$ will not contribute. Outside the derivative in
Eq.~(\ref{eqn:habg}), we need only the first term of
Eq.~(\ref{eqn:Iv2}) and we can replace $A'$ with $\bA'$. Thus we define
\begin{equation}\label{eqn:g}
    g(u,v) = \frac{s_{\alpha\beta}(u,v)X_\gamma}{A'_u(v)+u(\bB''\cdot A')/4}
\end{equation}
and using $A'_u = Z/2$ we can rewrite Eq.~(\ref{eqn:habg}) as
\begin{equation}\label{eqn:habg-generic}
    h_{\alpha\beta,\gamma} = \frac{8G\mu}{Z}\int\frac{du}{u^2} \left(\frac{\partial g_{\alpha\beta\gamma}}{\partial v}\right)\,.
\end{equation}
Because we would like to find contributions up to $\order(u)$ in the integrand, we will expand
\begin{equation}\label{eqn:gexp}
    \frac{\partial g}{\partial v} = \bg_{,v}+u\bg_{,uv}+\frac{u^2\bg_{,uuv}}{2}+\frac{u^3\bg_{,uuuv}}{6}+v\bg_{,vv}\,.
\end{equation}
We will not need higher orders.

Now $\bX = 0$, so $X_\gamma$ must be differentiated.  Furthermore,
$X_{u,u} = \bB'_u/2 =0$ and $X_{u,uu} = \bB''_u/2 =0$.  Thus in order to
have a $u$ component in some differentiated $X$, we need to
differentiate with respect to $v$, or 3 times with respect to $u$, and
vice versa.

On the other hand, $\bs_{u\beta} =\Sab(\bA',\bB') \bB'^\alpha = 0$,
so $\sab$ must be differentiated.  Differentiating with respect to $v$
just differentiates $A'$, so $\bs_{u\beta,v\ldots v}=0$ regardless of
the number of derivatives.  In order to have a $u$ component in $\bs$,
we much differentiate with respect to $u$, and the same for $v$.

Furthermore, $\bs_{uu,u}= 0$ because of Eq.~(\ref{eqn:Sigma-features}).
Additional derivatives with respect to $v$ make no difference.

Now let us find the leading order term in Eq.~(\ref{eqn:gexp}).  We
need two derivatives, one for $\sab$ and one for $X_\gamma$.  But
among $\alpha,\beta,\gamma$ there must be $u$ and $v$.  By the
considerations above, we thus need to differentiate $\sab$ and
$X_\gamma$ both with respect to $v$ or both with respect to $u$.  Thus
$\bg_{,v}$ and $\bg_{,uv}$ do not contribute, and the integral in
Eq.~(\ref{eqn:habg}) never diverges.

To go beyond this level, we need to consider the specific combinations
of indices we need in Eq.~(\ref{eqn:x}).  First consider $h_{vv,u}$.
This involves $s_{vv}$.  To get a term in Eq.~(\ref{eqn:gexp})
that doesn't vanish we need to go up to $s_{vv,vv}$.  Thus we
need the last term in Eq.~(\ref{eqn:gexp}), but both derivatives have
been applied to $s$, leaving none for $X$, so $h_{vv,u}= 0$ at
this order.

Now consider $h_{uu,v}$.  Here we need to differentiate $s$ twice
and $X$ once with respect to $u$.  Thus we take the penultimate term of
Eq.~(\ref{eqn:gexp}).  There's one derivative with respect to $v$ left,
and it acts on
\begin{equation}\label{eqn:suuuuA}
\frac{s_{uu,uu}}{A'_u}=\frac{2A'_uB'''_u}{A'_u} = 2\bB'''_u\,,
\end{equation}
which has no $v$ dependence, so $h_{uu,v} = 0$.

So we are interested now only in $X^c_{,uv}$ and $X^d_{,uv}$.  These
have exactly the same form, so we will compute only the former.

There are 3 terms with the indices in different orders.  First
consider $h_{vc,u}$.  To keep $s_{vc}$ from vanishing we need to
differentiate with respect to $v$.  Then we need to differentiate $X$
once with respect to $v$ or thrice with respect to $u$, using all the
rest of the derivatives in either case.  In the former case,
\begin{equation}
\bg_{,vv} = \Sigma_{vc}(\bA'',\bB')=\frac{\bA''_cZ}2
\end{equation}
Differentiating $X_u$ with respect to $v$ gave $A'_u/2$, canceling
the $A'_u$ in the denominator and a combinatoric factor of 2 from the
placement of the derivatives.
The other possibility gives
\begin{equation}
\bg_{,uuuv} = \frac{\Sigma_{vc}(\bA'',\bB')\bB'''_u}{2\bA'_u}=
-\frac{\bA''_c\bB''^2}4\,.
\end{equation}
These terms give a contribution from each $u$ to $X^c_{,uv}$ of
\begin{equation}\label{eqn:x2}
\frac{G\mu\bA''_c\bB''^2 u}{12Z}\,.
\end{equation}

Now we consider $h_{uc,v}$ and $h_{uv,c}$ together.  We'll need to
differentiate $s$ with respect to $u$, so $\bg_{vv}$ does not
contribute here.  The other terms have one $v$ derivative.  If we
apply it to $X_\gamma$, we get $\bA'_c=0$ or $\bA'_v= 0$, so we can
take $B_\gamma/2$ for $X_\gamma$.

Thus we take
\begin{equation}\label{eqn:hucv}
\frac{s_{uv} B_c- s_{uc} B_v}{2 (A'_u+ u\bB''\cdot A'/4)}\,,
\end{equation}
differentiate with respect to $u$ 2 or 3 times, set $u = 0$, and
differentiate with respect to $v$.

In the first term in the numerator, one derivative must act on
$s$, and two on $B_c$ giving
\begin{equation}\label{eqn:hucv1}
\frac{3s_{uv,u} \bB_c''}{2A'_u} = \frac{3(A'_u\bB''_v -(Z/4) 
A'\cdot\bB'')\bB_c''}{2A'_u}
\end{equation}
The first term has no $v$ dependence.

In the other term from Eq.~(\ref{eqn:hucv}), we need one derivative on
$s$, and one on $B_v$.  If we differentiate neither the denominator
nor $s$ (again), the only possible $v$ dependence is in
$s_{uc,u}/A'_u$, but this is just $B''_c$, because $\bB''_u= 0$.
So in these cases there's nothing to differentiate with respect to
$v$.

The remaining terms are
\begin{equation}\label{eqn:hucv3}
\frac{3s_{uc,uu}Z}{4A'_u}
-\frac{3s_{uc,u}(\bB''\cdot A')Z}{8(A'_u)^2}\,.
\end{equation}
The second term is
\begin{equation}\label{eqn:cancellation1}
\frac{3\bB''_c(\bB''\cdot A')Z}{8A'_u}
\end{equation}
and it cancels the second term in Eq.~(\ref{eqn:hucv1}).  We do not
know any good explanation for this cancellation.

The first term in Eq.~(\ref{eqn:hucv3}) is
\begin{equation}
\frac{3A'_c\bB'''_u Z}{4A'_u} = -\frac{3A'_c\bB''^2 Z}{8_u}
\end{equation}
plus a term with no $v$ dependence.  We must apply the $v$ derivative
to $\bA'_c$, so the contribution from $h_{uc,v}$ and $h_{uv,c}$ is
\begin{equation}
\bg_{,uuuv} =\frac{3A''_c\bB''^2}{4}\,,
\end{equation}
and the contribution to $X^c_{,uv}$ is
\begin{equation}\label{eqn:x3}
\frac{G\mu\bA''_c\bB''^2 u}{2Z}\,.
\end{equation}

Putting together Eqs.~(\ref{eqn:x2},\ref{eqn:x3}) gives
the total contribution to $X^c_{,uv}$ from a sufficiently close source point,
\begin{equation}\label{eqn:smooth-total}
\frac{7G\mu\bA''_c\bB''^2u}{12Z}
\end{equation}
The $d$ term is just the same, while from above $X^u_{,uv}=X^v_{,uv}= 0$.
One can write a total contribution from all sources nearer than
some small distance $\umax$,
\begin{equation}\label{eqn:smooth-int}
X^c_{,uv}=\frac{7G\mu\bA''_c\bB''^2}{12Z}\int_{-\umax}^0 u~du
= \frac{7G\mu\bA''_c\bB''^2}{12Z}\left(\frac{\umax^2}{2}\right)\,.
\end{equation}
Equation~(\ref{eqn:smooth-int}) reproduces the result of Appendix A in
Ref.~\cite{Quashnock:1990wv}.  But since this effect grows as we get
further from the observer, the total effect is dominated by distant places
where this calculation does not apply.

The main importance of this result is that there is no divergent
contribution from nearby points on a smooth worldsheet.  When there are
points where the worldsheet is not smooth, such as kinks and cusps,
this result does not apply and the effect may diverge as one approaches
these special points, as we now discuss.

\section{Close to a kink}\label{sec:kink}

We begin by introducing a kink in $A$ at $v=0$. We will take the
Taylor expansion of $B$ as before, but $A$ is no longer analytic, so
let us instead consider a form which is straight on each side of
the kink,
\begin{equation}
A(v) =\begin{cases}A'_- v & v< 0\\A'_+v & v > 0\end{cases}
\end{equation}
Curved segments of $A$ would not affect the divergent behavior.

\begin{figure}
    \centering
    \includegraphics[scale=0.75]{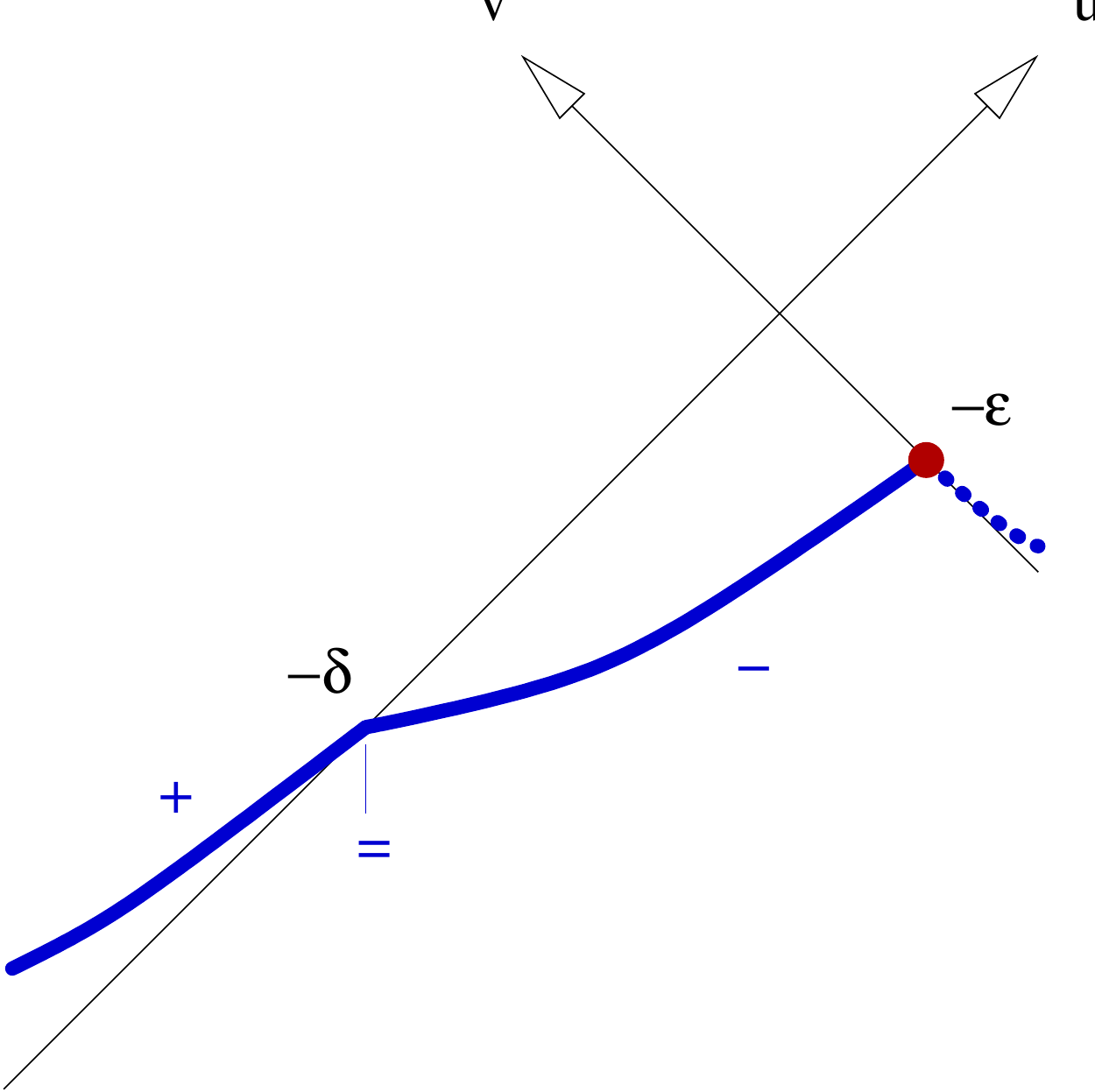}
    \caption{A drawing of an observer point (red circle) at
      $(u,v)=(0,-\epsilon)$ near a kink in $A$ at $v=0$. The
      intersection of the lightcone with the worldsheet is in blue,
      with the relevant branch solid and the other branch dotted.
      The region ``below'' the kink is labeled by $-$, above the
      kink by $+$, and at the kink by $=$. As the observer
      approaches the kink ($\epsilon\rightarrow0$), the distance at
      which the intersection crosses the kink, $\delta$, will also
      go to zero.}\label{fig:kink-drawing}
\end{figure}

We will consider our observer to be at $\bu=0$ and $\bv=-\epsilon<0$.
We will consider observers at $v > 0$ in Sec.~\ref{sec:abovekink}.
The past lightcone in the mostly negative $v$ direction does not
intersect the kink, so effect from such sources is the smooth result
of the previous section.  In the mostly negative $u$ direction it
intersects the kink at some point we will call $u=-\delta$. The
integral of Eq.~(\ref{eqn:habg}) therefore covers three regimes: when
$v<0$ and $u>-\delta$, which we call \emph{below} the kink, and denote
related quantities with a subscript or superscript $-$; when $v>0$ and
$u<-\delta$, which we call \emph{above} the kink, and which has
subscript or superscript $+$; and finally when $v=0$ and $u=-\delta$,
which we call \emph{at} the kink, and indicate by a subscript or
superscript $=$. Fig.~\ref{fig:kink-drawing}. shows an observer point
and these three regions of its intersection line.

In the region below the kink, the existence of the kink has no effect, and
the result is as in Sec.~\ref{sec:generic}, with no divergence.  When
the sources are above the kink, source quantities may no longer be
similar to quantities at the observer, as assumed in
Sec.~\ref{sec:generic}, so the calculation there are no longer applies
and divergences are possible.  In addition, at $u=-\delta$ there is a
discontinuous change in both $s_{\alpha\beta}$ and $\intervalv$.  Thus
the integrand in Eq.~(\ref{eqn:habg}) is a $\delta$-function in $u$,
leading also to a divergent effect.

Before considering the regions individually, we wish to determine the
relationship between $\delta$ and $\epsilon$. Let us start
at the observer and move backwards along the lightcone, primarily in
the $-u$ direction. We move first through the region below the kink,
where $A=vA'_-$, and so to lowest order in $v$ we find
\begin{subequations} \begin{align}
    (\dA)^2 &= \order(\epsilon^4)\,,\\
    (\dB)^2 &= -\frac{u^4\bB''^2}{12}\,,\\
    \dA\cdot\dB &= u(\epsilon+v)Z_-\,,
\end{align} \end{subequations} 
where $Z_\pm = \bB\cdot A'_\pm$, and thus $Z_-$ is the $Z$ of Eq.~(\ref{eqn:eta-uvcd}). Thus
\begin{equation}\label{eqn:ell-}
    \interval_- = \frac{(\epsilon+v)uZ_-}{2}-\frac{u^4\bB''^2}{48}\,.
\end{equation}
With the lightcone constraint $\interval=0$, this means that when we are at the kink and $(u,v)=(-\delta,0)$, we find
\begin{equation}\label{eqn:delta}
    \delta = \left(-\frac{24Z_-\epsilon}{\bB''^2}\right)^{1/3}\,.
\end{equation}

We next continue to the region above the kink. Now, $\bA-A = vA'_+ + \epsilon A'_-$, and so, ignoring terms like $v^2$ and $u^4$ or higher, we find
\begin{equation}\label{eqn:ell+}
    \interval_+ = \frac{\epsilon u Z_- + vu Z_+}{2}-\frac{u^4\bB''^2}{48}\,.
\end{equation}
This allows us to write the general relationship
\begin{equation}
    v(u)_\pm = \frac{(u^3+\delta^3)\bB''^2}{24Z_\pm}\,.
\end{equation}
Since we're concerned with $u$ of order $\delta$, $v(u)$ is of order
$\delta^3$, and so we will be concerned only with terms at most
linear in $v$.

Before moving on, we note that we can also write the general relationship
\begin{equation}
    \intervalv^\pm = \frac{A'_\pm\cdot B}{2} = uA'_{u\pm} + \frac{u^2A'_\pm\cdot\bB''}{4}\,,
\end{equation}
which is necessary for finding the denominator of the acceleration integrand.

Now, we can consider how divergences might arise as we integrate along
the intersection line with respect to $u$, starting above the kink and
crossing it.

\subsection{Divergent behavior above the kink}\label{ssec:divergent+}

We begin on the side of the kink with $v>0$, $u<-\delta$. Here, the
only thing in Eq.~(\ref{eqn:habg}) that can be differentiated with
respect to $v$ is $\dX_\gamma$, and so find to lowest order that
\begin{equation}
  h^+_{\alpha\beta,\gamma} =
    \frac{8G\mu}{Z^2_+}\int^{-\delta}du\frac{\bs^+_{\alpha\beta}A'_{+\gamma}}{u^2}
    = \frac{8G\mu \bs^+_{\alpha\beta}A'_{+\gamma}}{Z^2_+\delta}\,.
\end{equation}
We have included only the upper limit of integration, which would be
the source of terms that diverge for small $\delta$.  If we expand
to one more order in $u$, we expect divergences of order
$\ln\delta$, but we will not attempt to compute those.

Consulting Eq.~(\ref{eqn:x}), we see that all terms involve at least
one $u$ index. But $\bs_{u\beta}^+=0$ from
Eq.~(\ref{eqn:Sigma-features}). Thus we must have $\gamma=u$, and so
the only metric perturbation terms we need to consider
are 
\begin{subequations} \begin{align} h^+_{vv,u} &=
  \frac{2G\mu(A'_+\cdot
    A'_-)Z_-}{Z_+\delta}\,,\label{eqn:h+vvu}\\ h^+_{vc,u} &=
  \frac{2G\mu A'_{+c}Z_-}{Z_+\delta}\,.\label{eqn:h+vcu}
\end{align} \end{subequations}
The terms with $d$ instead of $c$ are analogous.

\subsection{Divergent behavior at the kink}\label{ssec:divergent=}

Now we consider divergences as we integrate across the kink, where
$u= -\delta$, $v = 0$.  There is no jump in $\dX$ there, but
$\sab$ and $\intervalv$ change discontinuously.  So we 
define $F^+_{\alpha\beta}$ to be the value of $\sab/\intervalv$
immediately above the kink and $F^-_{\alpha\beta}$ to be the value
immediately below,
\begin{equation} \label{eqn:F}
    F^\pm_{\alpha\beta} = -\frac{s^\pm_{\alpha\beta}}{\delta Z_\pm + \delta^2(A'_\pm\cdot\bB')/4} = -\left(\frac{2}{\delta Z_\pm}-\frac{A'_\pm\cdot\bB''}{Z^2_\pm}\right)s^\pm_{\alpha\beta}\,,
\end{equation}
plus higher orders in $\delta$.  For most of our purposes, we will only
need the $1/\delta$ term, but the latter will be important later
on. Now, we write
\begin{equation}
    h^=_{\alpha\beta,\gamma} = 2G\mu\int du\frac{\delta(v)(F^+-F^-)_{\alpha\beta}(\epsilon A'_--B(-\delta))_\gamma}{\intervalv}\,.
\end{equation}
We now substitute $v(u)$ given by $\interval=0$ and use the relation
\begin{equation}
    \frac{\delta(v)}{\intervalv} = \frac{\delta(u+\delta)}{\intervalu}\,.
\end{equation}
Now $\intervalu =\dX\cdot B'$, and at the kink crossing this becomes
\begin{equation}
\intervalu = \frac{\epsilon Z_-}{2} + \frac{\delta^3 \bB''^2}{12}
= \frac{\delta^3 \bB''^2}{16}\,,
\end{equation}
so
\begin{equation}
    h^=_{\alpha\beta,\gamma} = \frac{32G\mu(F^+-F^-)_{\alpha\beta}(\epsilon A'_--B(-\delta))_\gamma}{\delta^3\bB''^2}\,.
\end{equation}

We will now consider specific indices of the metric perturbation derivatives in order to find the divergent behavior of the accelerations.

\subsubsection{Divergences for $\gamma=u$}\label{sssec:u}

First, consider $\gamma=u$ and expand $\bB(-\delta)$. The first nonvanishing term is $\delta^3\bB''^2/12$, which combines with $\epsilon A'_{-u}$ to give $\delta^3\bB''^2/16$, and so
\begin{equation}
    h^=_{\alpha\beta,u} = 2G\mu(F^+-F^-)_{\alpha\beta}\,.
\end{equation}
We are interested only in $\alpha\beta=vv$ and $\alpha\beta=vc$.
When we choose $vv$, $F^-=0$ and
\begin{equation}
    F^+_{vv}=-\frac{(A'_+\cdot A'_-)Z_-}{Z_+\delta}
\end{equation}
to leading order, and thus
\begin{equation}\label{eqn:h=vvu}
    h^=_{vv,u} = -\frac{2G\mu(A'_+\cdot A'_-)Z_-}{Z_+\delta}\,.
\end{equation}
This cancels the term in Eq.~(\ref{eqn:h+vvu}). We have calculated all
possibly divergent components of the $u$ direction acceleration, and
as a consequence of this cancellation we find that $X^u_{,uv}$ has
no $1/\delta$ divergence.

When we choose $vc$, we again have $F_-=0$, but now to leading order we find
\begin{equation}
    F^+_{vc} = -\frac{A'_{+c}Z_-}{Z_+\delta}\,,
\end{equation}
and therefore
\begin{equation}\label{eqn:h=vcu}
    h^=_{vc,u} = -\frac{2G\mu A'_{+c}Z_-}{Z_+\delta}\,.
\end{equation}
Once again, this cancels the above-kink region contribution, and so
terms like $h_{vc,u}$ are not divergent.  The reason for these
cancellations can be seen by rewriting Eq.~(\ref{eqn:habg}) using
$\partial/\partial u$ instead of $\partial/\partial v$.

\subsubsection{Divergences for $\gamma=v$}\label{sssec:v}

Now we consider terms with $\gamma=v$. Because $A'_{-v}=0$, we need $B_v=\delta Z_-/2$, and therefore
\begin{equation}
    h^=_{\alpha\beta,v} = -\frac{16G\mu Z_-(F^+-F^-)_{\alpha\beta}}{\bB''^2\delta^2}\,.
\end{equation}
The only two choices of $\alpha\beta$ we need to consider are $uu$ and
$uc$.  For the former,
\begin{equation}
    F^\pm_{uu} = \bB''^2\delta
\end{equation}
to first order, so $F^+_{uu} = F^-_{uu}$.  Thus $h^=_{uu,v}=0$, so $X^v_{,uv}$ has
no $1/\delta$ divergence.

Now consider $uc$. Here we must take into account both terms of Eq.~(\ref{eqn:F}). Moreover, we will consider the two terms in
\begin{equation}
    s^\pm_{uc} = A'_{\pm u}B'_c + A'_{\pm c} B'_u
\end{equation}
individually.

Starting with the $A'_{\pm u}B'_c$ term, and with $B'_c = -\delta\bB''_c$ when $u=-\delta$, we find that for this term
\begin{equation}
    F^\pm_{uc} = \left(1+\frac{(A'_\pm\cdot\bB'')\delta}{2Z_\pm}\right)\bB''_c
\end{equation}
and therefore a contribution to the metric perturbation of
\begin{equation}
    \frac{8G\mu Z_-\bB''_c}{\delta\bB''^2}\left[\frac{A'_+\cdot\bB''}{Z_+}-\frac{A'_-\cdot\bB''}{Z_-}\right]=\frac{8G\mu Z_-\bB''_c}{\delta\bB''^2}(A'_+\star\bB'')\,.
\end{equation}
Then taking the $A'_{\pm c}B'_u$ term, we must go to $B'_u = \delta^2\bB'''_u/2 = -\delta^2\bB''^2/2$. Thus, for this term,
\begin{equation}
    F^+_{uc} = \frac{\delta A'_{+c}\bB''^2}{Z_+}\,.
\end{equation}
Of course, $A'_{-c}=0$, and so $F^-_{uc}=0$ for this term. So in sum,
\begin{equation}\label{eqn:h=ucv}
    h^=_{uc,v} = \frac{8G\mu Z_-(A'_+\star\bB'')\bB''_c}{\delta\bB''^2} - \frac{8G\mu Z_- A'_{+c}}{\delta Z_+}\,.
\end{equation}

\subsubsection{Divergences for $\gamma=c$}\label{sssec:c}

The remaining choice for $\gamma$ is $c$. Now the leading term comes from $B_c=-\delta^2\bB''_c/2$, giving
\begin{equation}
    h^=_{\alpha\beta,c} = \frac{16G\mu(F^+-F^-)_{\alpha\beta}\bB''_c}{\delta\bB''^2}\,.
\end{equation}
But now, the only choice for $\alpha\beta$ that we can make is $uv$. At leading order,
\begin{equation}
    s^+_{uv} = -\frac{(A'_+\star B')Z_-}{4} = \frac{\delta(A'_+\star\bB'')Z_-}{4}
\end{equation}
and $s^-_{uv}=0$, and thus $F^-_{uv}=0$ as well. Thus
\begin{equation}
    h^=_{uv,c} = \frac{8G\mu Z_-(A'_+\star\bB'')\bB''_c}{\delta\bB''^2Z_+}\,.
\end{equation}
This is identical to the first term of Eq.~(\ref{eqn:h=ucv}), and
contributes oppositely in Eq.~(\ref{eqn:x}).  This cancellation is
analogous to the one involving Eq.~(\ref{eqn:cancellation1}). The only
remaining $1/\delta$ divergent term for the $c$ direction acceleration is the
second half of Eq.~(\ref{eqn:h=ucv}), giving
\begin{equation}\label{eqn:acc-kink-c}
    X^c_{,uv} = \frac{4G\mu Z_- A'_{+c}}{\delta Z_+}
= -\frac{2G\mu A'_{+c}}{Z_+}\left(\frac{\bB''^2Z_-^2}{3\epsilon}\right)^{1/3}\,.
\end{equation}
Thus the transverse accelerations diverge as an observer
approaches a kink, but only as the inverse cube root of the
distance.  Equation~(\ref{eqn:acc-kink-c}) agrees with the
acceleration reported in Ref.~\cite{0264-9381-22-13-002} for the loop
discussed there.

\subsection{Observers above the kink}\label{sec:abovekink}

In the previous subsections, we considered observers below the kink,
i.e., points that the kink is approaching.  Here we will show that
there are no divergences for observation points above the kink, i.e.,
where the kink has already passed by.  We keep the forms of $A$ and
$B$ above, but now we consider an observation point with $\bu = 0$,
$\bv = \epsilon > 0$.  The backward lightcone that intersects the kink
is the one mostly in the negative $v$ direction.  The intersection
occurs at a point $v=0$, $u =\delta >0$, with $\delta = \order(v^3)$.
This is the critical difference: because the lightcone now starts in the
$-v$ direction, perpendicular to the kink motion, it quickly reaches the
kink with little transverse motion.

We will use the $u$-$v$ exchanged version of Eq.~(\ref{eqn:habg}),
\begin{equation}\label{eqn:habgu}
    h_{\alpha\beta,\gamma} (\bX)=4G\mu\int dv \left[\frac{1}{\intervalu}\,\frac{\partial}{\partial u}\left(\frac{s_{\alpha\beta}\dX_\gamma}{\intervalu}\right)\right]_{u=u(v)}\,.
\end{equation}
Applying $\partial/\partial u$ does not lead to any
$\delta$-functions, because the $u$ direction does not cross the kink.

Now
\begin{equation}
\intervalu = \dX\cdot B' = (A_\pm v - A_-\epsilon)/2\cdot B'
=(Z_\pm v - Z_-\epsilon)/2\,,
\end{equation}
where we ignore $\order(\delta)$.  We will be concerned with $v$ of
order $\epsilon$, in which case $\intervalu = \order(\epsilon)$, and
$\intervalu$ does not vanish as $v\to 0$.  (It does vanish as
$v\to\epsilon$, but this is just the near-observer regime of
Sec.~\ref{sec:generic}.)  Furthermore, $\intervalu$ has no $u$
dependence.  Thus in Eq.~(\ref{eqn:habgu}) we must differentiate
either $\sab$ or $\dX_\gamma$.  In the former case, we are left with
$\dX_\gamma = O(\epsilon)$.  Thus the integrand is
$\order(\epsilon^{-1})$, and since the range of integration is
$\order(\epsilon)$, the result is at most a constant in $\epsilon$.

The other possibility is that we apply $\partial/\partial u$ to
$\dX_\gamma$, giving $B'_\gamma/2$, and leave $\sab$ undifferentiated.
Since we are ignoring $\order(\delta)$, we can take $B'$ as $\bB'$
both in $\dX_\gamma$ and in $\sab$.  But $\bB'$ has only one nonzero
component, which is $v$.  Thus $\gamma$ must be $v$ and also one of
$\alpha$ and $\beta$ must be $v$ (or both must be $c$ or $d$), but no
such term appears in Eq.~(\ref{eqn:x}).  Thus there's no divergence
for observers approaching the kink from above.

\subsection{Changes to the string near a kink}\label{ssec:kink-changes}

What does Eq.~(\ref{eqn:acc-kink-c}) tell us about how the worldsheet
is modified around a kink?  Because the kink we studied is at a fixed
position in $v$, the effects on $A'$ and $B'$ are different.  To find
the correction to $B'$ at a certain fixed $u$, we integrate around the
worldsheet in the $v$ direction, following
Eq.~(\ref{eqn:delta-nv}). This line of integration will always pass
across the kink and, since the divergent part of the acceleration near
the kink is only like $v^{-1/3}$, there is no divergence after
integration with respect to $v$.  In fact, as discussed in
Sec.~\ref{sec:worldsheet}, since no divergence appears in $\dB'$, we
cannot say for sure that there is a divergent effect on $B'$ at all.

Conversely, we find the correction to $A'$ by fixing $v$ and
integrating around the worldsheet in the $u$ direction. The kink
always remains the same distance away, and the divergent $v^{-1/3}$
behavior remains in the correction to $A'$. This correction is always
transverse to the worldsheet, but the worldsheet direction changes as
we integrate the corrections to $A'$ at different observation points.
Thus the divergent correction to $A'$ for a whole oscillation is quite
general, except that it must be perpendicular to $A'$, so that $A'$
remains null.  This divergence cannot be a gauge artifact.

The loss of length of the string is given by the change to the time
component of $A'$, which generally diverges as $\bv^{-1/3}$.  The
total loss of length gives the total energy emitted from the string.
To compute this we integrate over $\bv$, which gives a finite result
as it should \cite{Garfinkle:1987yw}.

Now we will estimate the length scale at which a kink is rounded
off. Define
\begin{equation}
    K^\gamma=A'^\gamma_+-A'^\gamma_-
\end{equation}
for the tangent vectors at a pair of points of fixed $v$ above and
below the kink. This is the kink's ``turning vector'' across that
range in $v$, so decreases in $K$ constitute smoothing the kink out to
that range.  We will assume that the back-reaction is not affected by
smoothing closer to the kink than the points of interest, so we can
use Eq.~(\ref{eqn:acc-kink-c}), which we rewrite as
\begin{equation}\label{eqn:acc-kink-c-alt}
   X^c_{,uv} = -\frac{2 G\mu}{Z_+}
   \left(\frac{\bB''^2Z_-^2}{3L}\right)^{1/3}\left(\frac
   Lv\right)^{1/3}A'_{+c}\,.
\end{equation}
This modifies the vector $A'_-$, making it closer (because $Z_+ < 0$)
to $A'_+$, and so decreasing the bending angle.  However, the change in
$A'_-$ is given by the projection of $K$ into directions transverse to
the worldsheet,
\begin{equation}
    K_\perp = A'_+ - \frac{YB'+Z_+A'_-}{Z_-}\,,
\end{equation}
with $Y=A'_+\cdot A'_-$.  The length of $K$ will be modified according to
how much the transverse acceleration points in the direction of $K$,
i.e., the magnitude of $K_\perp\cdot K/|K|$, introducing an overall factor
\begin{equation}
    \frac{K_\perp\cdot K}{K^2} = \frac{-YZ_+/Z_--YZ_+/Z_--Y+Y}{-2Y} = \frac{Z_+}{Z_-}\,,
\end{equation}
which may be more or less than one because of the Lorentzian metric.
The instantaneous change to the length of $K$ at a particular point is thus
\begin{equation}
    |K|' = -4G\mu\left(\frac{\bB''^2}{3LZ_-}\right)^{1/3}\left(\frac Lv\right)^{1/3}|K|\,,
\end{equation}
where there is a factor of 2 from Eq.~(\ref{eqn:delta-nv}).

Now we integrate this projection with respect to $\bu$ over one
oscillation. This tells us about the rate of change of the length of
$K$ per oscillation. Dividing by the loop oscillation time of $L/2$ 
converts this to an average rate of change,
\begin{equation}
    \frac{d|K|}{dt} = -\frac{G\mu H}{L}\left(\frac{L}{v}\right)^{1/3}|K|\,,
\end{equation}
where the dimensionless coefficient is given by
\begin{equation}
H = \int_0^L d\bu ~8\left(\frac{\bB''^2}{3LZ_-}\right)^{1/3}\,.
\end{equation}

Thus, $|K|$ decreases exponentially with time, with a time constant of
$(G\mu H/L)(L/v)^{1/3}$, so the kink has been significantly rounded
off to distance $v$ after a time
\begin{equation}
    t_{\text{kink}} \approx \frac{L}{G\mu H}\left(\frac vL\right)^{1/3}\,.
\end{equation}

\begin{figure}
    \centering
    \includegraphics[scale=1.00]{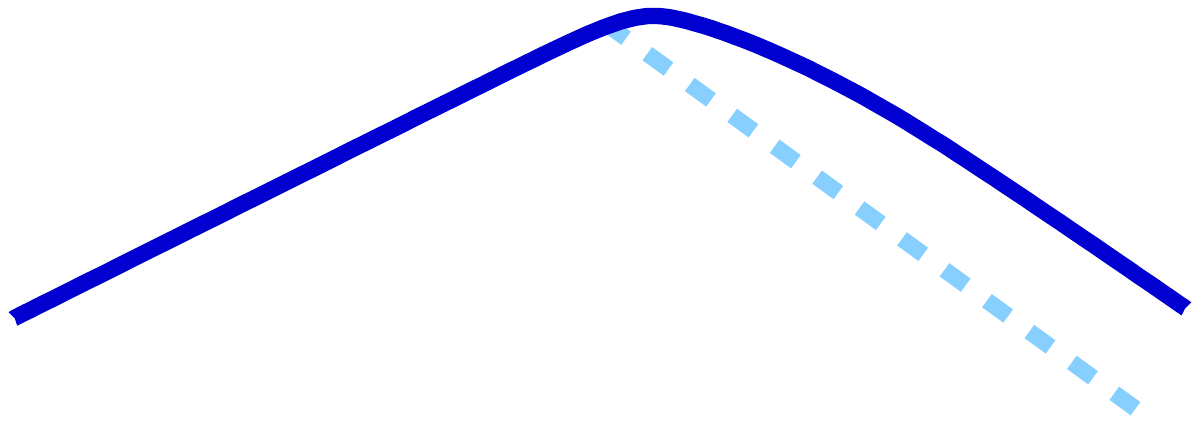}
    \caption{How a kink is modified due to back-reaction. We show a
      segment of a worldsheet function, where the region above the
      kink (solid blue, on the left) does not change, but the region
      below the kink goes from being straight (dashed light blue) to
      having some curvature (solid blue, on the right). Note that the
      curvature dies out as one goes to the right, so there is some
      distance after which the $A'$ below the kink before and after
      back-reaction are effectively identical.}\label{fig:a-change}
\end{figure}

The loop's lifetime is $t\approx L/(\Gamma G\mu)$, with $\Gamma$ the
measure of the loop's power loss rate.  At the end of the loop's
lifetime, we can estimate that significant rounding extends to a
distance
\begin{equation}
    v_{\text{rounded}} \approx \left(\frac{H}{\Gamma}\right)^3L\,.
\end{equation}
We show a drawing of this rounding process in Fig.~\ref{fig:a-change}.

Because $\Gamma$ is of order $50$ for realistic loops, the rounding
distance may be much less than $L$. Let us consider
a ``generic'' loop, which has worldsheet functions which are mostly
smooth circles except for a few large kinks.  We take as typical
values $|B''| = 2 \pi/ L$, $Z_\pm = -1$, so $H=8(4\pi^2/3)^{1/3}\approx
20$ and $(H/\Gamma)^3\approx 0.06$.  This means that rounding process
never has much effect on regions further from the kink than about
$0.06 L$; at such distances the kink mostly retains its original appearance.

If the kink is preventing the occurrence of a cusp, by jumping over
what would otherwise be an intersection between $A'$ and $B'$, the
smoothing process will reintroduce the cusp.  However, the cusp will
be weak, in the sense that little of the total string length will ever
be involved in it.  Of course this is a very simplified model.
Strings taken from simulations have many kinks of various angles, and
little smoothly bending parts of the string, so this analysis does not
apply.

Our estimate of how the kink is rounded is only good if the change in
one oscillation is small. This means that we require
$v/L>(4G\mu H)^3$, but this is an incredibly tiny number, and so the
preceding is valid until we are extremely close to the kink. For
example, using roughly the current observational upper bound of
$G\mu=10^{-11}$ and our estimate of $H$ above, we find $v/L\gtrsim
10^{-30}$ as our requirement.

\section{Close to a cusp}\label{sec:cusp}

Now, we consider an observation point on a string with smooth $A$ and
$B$, but place the observer very near to a cusp. As mentioned in
Sec.~\ref{sec:setup}, a cusp is formed when $\mathbf{A}'=\mathbf{B}'$
or equivalently $A'=B'$, so points near a cusp have $Z = A' \cdot B'
\ll1$.  Otherwise-well-behaved quantities such as
Eq.~(\ref{eqn:smooth-total}) may thus diverge as the observation point
approaches a cusp.  We now analyze this situation.

\subsection{Coordinate system}\label{ssec:wmpq}

While the $uvcd$ coordinates greatly simplified our investigations of
the kink (and the generic point), they are not well adapted to
studying the cusp.  If we define the $uvcd$ basis at a point near the
cusp, the vanishing of $Z$ leads to divergences in the metric and the
lengths of the basis vectors, which make it difficult to distinguish
actual divergences from coordinate divergences.  Instead we will use a
fixed basis for all points near the cusp, which we now define.

Let $e_{(w)} = A'/2$ (equivalently, $B'/2$) at the cusp, and let
$e_{(m)}$ be $w$ with its spatial component reversed.  Then let
$e_{(p)}$ and $e_{(q)}$ be any unit spacelike vectors orthogonal to
$e_{(w)}$, $e_{(m)}$, and to each other.  In the $wmpq$ basis, the
metric tensor is
\begin{equation}\label{eqn:eta-wmpq}
    \eta_{\alpha\beta}  = \left(\begin{array}{cccc} 0 & -1/2  & 0 & 0\\ -1/2 & 0 & 0 & 0\\ 0 & 0 & 1 & 0\\ 0 & 0 & 0 & 1\end{array}\right)\,,\qquad\eta^{\alpha\beta}  = \left(\begin{array}{cccc} 0 & -2 & 0 & 0\\ -2 & 0 & 0 & 0\\ 0 & 0 & 1 & 0\\ 0 & 0 & 0 & 1\end{array}\right)\,,
\end{equation}

Like its $uvcd$ cousin, this basis allows some simplifications in
vector components. We expand about the cusp as in
Eq.~(\ref{eqn:taylor}), getting
\begin{subequations}\label{eqn:wmpq-order}\begin{align}
    A_w &= -\frac{A''^2_0}{12}v^3\,,\\
    A_m &= -v+\frac{A''^2_0}{12}v^3\,,\\
    A_p &= \frac{A''_{0p}}{2}v^2\,,
\end{align} \end{subequations}
with $A_q$ just $A_p$ with $p\rightarrow q$, and the $B$ dependencies the same under $A\rightarrow B$, $v\rightarrow u$. Then we find $v$ or $u$ dependence for any derivative of $A$ or $B$ by applying the appropriate number of derivatives and taking the lowest-order term.

We now take the cusp to be at the origin, and the observer to be at
some point on the worldsheet $(\bu,\bv)$ near the cusp. When we
consider how different sources will affect the observer, we see that
there are two regimes: one for when the sources are much closer to the
observer than to the cusp, and one for when they are very far from
either the observer or the cusp.

In the former case, the sources do not know about the cusp, and
so the problem reduces to that of Sec.~\ref{sec:generic}, but the
resulting effect may be quite large because $Z\ll 1$, i.e., the string
is rapidly moving.  But when the sources are far from the observer
they cannot distinguish the observer from the cusp, and as a result
their contributions to the acceleration integrand grow divergently.

Because the scale at which this growth is cut off is when the source 
is about as far from the observer as the observer is from the cusp, we 
may see divergent accelerations as the observer moves towards the 
cusp. Let's find such an effect now by finding the general form of the 
acceleration integrand, and thereby the leading-order divergent term 
in the acceleration.

\subsection{Sources far from the observer}\label{ssec:far-sources}

Because we are now working with our origin at the cusp itself, we will
make the replacement $B'_0\rightarrow A'_0$ for the remainder of this
section. We can now also write
\begin{subequations} \begin{align} 
    A'_0\cdot B''_0 &= 0\,,\\
    A'_0\cdot B'''_0 &= -B''^2_0\,.
\end{align} \end{subequations}
We are considering sources close to the cusp, but much further from
the cusp than the observer is.  Thus we work in the regime $\bu,
\bv \ll u, v\ll L$.  Then the leading terms in $\interval$ are those
that which have a combined order in $u$ and $v$ of four, and the
lightcone constraint becomes
\begin{equation}\label{eqn:ellsq=0}
    0 = \interval =  \frac{\oAB}{8}u^2v^2-\frac{\oB''^2}{12}u^3v-\frac{\oA''^2}{12}uv^3-\frac{\oA''^2}{48}v^4-\frac{\oB''^2}{48}u^4\,.
\end{equation}
Solving this homogeneous quartic gives $v(u)=\lu u$, with $\lu$ some constant depending on the cusp parameters.

Rewriting Eq.~(\ref{eqn:habg}) as
\begin{equation}\label{eqn:habg-num-den}
    h_{\alpha\beta,\gamma}=-2G\mu\int du\frac{\intervalv\left[s_{\alpha\beta,v}(A+B)_\gamma+s_{\alpha\beta}A'_\gamma\right]-\intervalvv\left[s_{\alpha\beta}(A+B)_\gamma\right]}{\intervalv^3}
\end{equation} 
leads us to our next considerations: what are the lowest-order terms
in $u$ once we have contracted the $\bA'$ and $\bB'$ vectors into the
Christoffel symbol and made the replacement $v=v(u)$?
To lowest order in $u$ and $v$,
\begin{subequations} \begin{align}
    A\cdot\oA'' &= \frac{v^2}{2}\oA''^2\,,\\
    B\cdot\oA'' &= \frac{u^2}{2}(\oAB)\,,
\end{align} \end{subequations}
and the contractions with derivatives of $A$ and $B$ follow from there. From Eq.~(\ref{eqn:Sigma}), we can write
\begin{subequations}\label{eqn:sigma-a2b2}\begin{align}
    s_{\sigma\alpha}\oA''^\alpha &= (u\oAB)A'_{0\sigma}+(v\oA''^2)B'_{0\sigma}-(A'\cdot B')A''_{0\sigma}\,,\\
    s_{\sigma\alpha,v}\oA''^\alpha &= (u\oAB)A''_{0\sigma}+(\oA''^2)B'_{0\sigma}-(A''\cdot B')A''_{0\sigma}\,.
\end{align} \end{subequations}

As a final step before considering particular accelerations, we note that, to lowest order,
\begin{subequations}\label{eqn:habgs}\begin{align}
    h_{\beta\sigma,\alpha}\bA'^\alpha\bB'^\beta &= 4h_{\sigma w,w}\,,\\
    h_{\sigma\alpha,\beta}\bA'^\alpha\bB'^\beta &= 4h_{\sigma w,w}\,,\\
    h_{\beta\sigma,\alpha}\bA'^\alpha\bB'^\beta &= 4h_{ww,\sigma}\,.
\end{align} \end{subequations}
So now we have all the ingredients necessary to begin calculating the orders of the metric perturbation (thus acceleration) integrands. While the integrand numerators depend critically on the acceleration direction, the denominators are always the same. We will always write
\begin{equation}
    \intervalv^3|_{v=v(u)} = d_0 u^9
\end{equation}
where
\begin{equation}
    d_0 = \left[\frac{\lu\left(\oAB-\lu(\lu+3)\oA''^2\right)-\oB''^2}{12}\right]^3\,.
\end{equation}
The simple form of the denominator leads to the simple form of $d_0$. The numerator coefficients, which we will introduce in the following subsections, are generally far more complicated.

\subsection{$w$-direction acceleration}\label{ssec:acc-w-cusp}

We know that $g^{w\alpha}=0$ unless $\alpha=m$. Thus, in Eq.~(\ref{eqn:xuv}) with $\gamma=w$, it must be that $\rho=m$ everywhere. We turn to Eq.~(\ref{eqn:habgs}) to determine the orders of the terms involved.

Consider terms like $h_{ww,m}$ and $h_{wm,w}$. The terms in the numerator of Eq.~(\ref{eqn:habg-num-den}) are generally of three types. The first two are (where each vector has its own index) $A'B'A'$, $A'B'B'$, $A''B'A$, or $A''B'B$ multiplied by $\intervalv$; the third is $A'B'A$ or $A'B'B$ multiplied by $\partial^2\interval/\partial v^2$. Thus, based on Eq.~(\ref{eqn:wmpq-order}), we see that the lowest-order terms in the numerator are like $u^7$.

Thus, for accelerations in the $w$ direction, a source point $u$ away contributes
\begin{equation}\label{eqn:acc-w-cusp}
    G\mu~\frac{n_{0w}}{d_0} \left(\frac{1}{u^2}\right)\,,
\end{equation}
where $n_{0w}$ is, like $d_0$, a constant which depends on the cusp parameters. It is more complicated than $d_0$, owing to the greater complexity of the numerator:
\begin{align}
    n_{0w} &= \frac{\oA''^4(\oAB)}{144}\left(\lu^6+6\lu^5\right)-\frac{\oA''^2}{48}\left(\oA''^2\oB''^2+4(\oAB)^2\right)\lu^4+\frac{11\oA''^2\oB''^2(\oAB)}{144}\lu^3\nonumber\\
    &\quad+\frac{\oA''^2\oB''^2}{96}\left(\oAB-3\oB''^2\right)\lu^2-\frac{\oA''^2\oB''^4}{48}\lu+\frac{\oB''^2(\oAB)}{144}\,.
\end{align}

Because the integrand has a divergence like $1/u^2$, the acceleration has a divergence like the inverse distance from the observer to the cusp\footnote{And also like the logarithm of the same, if we continue to further orders.}.

\subsection{$m$-direction acceleration}\label{ssec:acc-m-cusp}

Here, we use the same property of $g^{\alpha\beta}$ as above, but now replace in Eq.~(\ref{eqn:xuv}) all $\gamma$ by $w$. This leads to a number of cancellations when combining the terms in Eq.~(\ref{eqn:habgs}), meaning that we need only consider $h_{ww,w}$ and terms where the second derivative vectors are contracted onto the $(A+B)$ or $A'$ in Eq.~(\ref{eqn:habg-num-den}).

Consulting the same equations as before, we see that these are perhaps the highest-order indices one could choose. The $h_{ww,w}$ has terms like $u^9$, and thus each source contributes
\begin{equation}\label{eqn:acc-m-cusp}
    G\mu\frac{n_{0m}}{d_0}\,.
\end{equation}
There are no divergences in this direction.

\subsection{$p$-direction acceleration}\label{ssec:acc-p-cusp}

Because the $p$ and $q$ directions are interchangeable, we only need to calculate the divergent behavior of one of them.

The only non-zero metric component involving $p$ is $g^{pp}$. Thus, for finding $X^p_{,uv}$, we set $\rho=p$ everywhere in Eq.~(\ref{eqn:xuv}). There are no cancellations.

We first consider terms like $h_{wp,w}$ and $h_{ww,p}$. They yield terms like $u^8$, and so the contribution for each source is
\begin{equation}\label{eqn:acc-p-cusp}
    G\mu ~\frac{n_{0p}}{d_0} \left(\frac{1}{u}\right)\,.
\end{equation}
This integrand has a divergence like $1/u$, and so the accelerations in the $p$ and $q$ directions diverge as the logarithm of the distance between the observer and the cusp.

\subsection{Total behavior of the cusp acceleration integrand}\label{ssec:cusp-integrand}

We now know how the acceleration for an observer near the cusp depends on the observer position for very distant sources. From Sec.~\ref{sec:generic}, we know that very near the observer, the integrand goes like $u$ only in the $c$ and $d$ directions. Now, we are interested to know how the cusp acceleration depends on the observer position when the sources are much closer to the observer than the observer is to the cusp, in order to compare the importance of the far and near regions of the integrand.

To do this, we express the contribution to the acceleration of a source point very near the observer as
\begin{equation}\label{eqn:acc-i-generic-vector}
\frac{7G\mu}{12}\frac{\bB''^2}{\bA'\cdot\bB'}\left[\bA''^\gamma-\frac{\bA''\cdot\bB'}{\bA'\cdot\bB'}\bA'^\gamma\right]=
    -\frac{7G\mu}{6}\frac{\oB''^2}{(\bv\oA''-\bu\oB'')^2}\left[\bA''^\gamma-\frac{2\oA''\cdot(\bv\oA''-\bu\oB'')}{(\bv\oA''-\bu\oB'')^2}\bA'^\gamma\right]u\,,
\end{equation}
which is nothing but the expression for a regular point,
Eq.~(\ref{eqn:smooth-total}), but now in
four-vector form. We see that it might be possible for the coefficient
to the $u$ to grow as $\bu,\bv\rightarrow0$, depending on the orders
of the components of $A'$ and $A''$. But finding the orders of those
components via Eq.~(\ref{eqn:wmpq-order}) shows that this will only be
a concern for the $w$ direction.

To show this, consider a line of worldsheet points lying in some specific
direction from the cusp, given by $\bv=\chi\bu$, with
$\chi$ some constant. Making this substitution and using
Eq.~(\ref{eqn:wmpq-order}) to find $A'$ and $A''$
components\footnote{Note that we now want the \emph{upper} index
  vectors, as opposed to the \emph{lower} index vectors as given in
  Eq.~(\ref{eqn:wmpq-order}), and so we use e.g. $P^w=\eta^{wm}P_m$.},
we find that the contribution per source in the $m$
direction goes as $u/\bu$, in the $p$ and $q$ directions goes as
$u/\bu^2$, and in the $w$ direction is
\begin{equation}\label{eqn:acc-i-w}
    \frac{7G\mu}{3}\frac{\chi\oA''\cdot(\chi\oA''-\oB'')\oB''^2}{(\chi\oA''-\oB'')^4}\frac{u}{\bu^3}\,.
\end{equation}
Upon integration of $u$ up to something proportional to $\bu$, the $m$,
$p$, and $q$ directions do not increase as $\bu\rightarrow0$. But,
something interesting has happened with the $w$ component. While the
integrand itself is linear in $u$ very near the observer, the
coefficient has a $1/\bu^3$ dependence. As a consequence, the
$w$-direction acceleration diverges as $1/\bu$ in the near regime,
just as it does in the far regime. Thus any estimate of the
acceleration for a point near a cusp must account for the effect of
both of those regimes.

Moreover, the signs of these effects do not need to be the same. For sources very far away, all observers near the cusp see contributions from such distant points as having the same sign, as $n_{0w}$ and $d_0$ are independent of $\bu$ and $\bv$. But consider Eq.~(\ref{eqn:acc-i-w}). Here, the overall sign depends on the sign of $\chi$, and the leading $1/\bu$ means that sign will always be different for two points with the same $\chi$ on opposite sides of the cusp.

\begin{figure}
    \centering
    \includegraphics[scale=1.0]{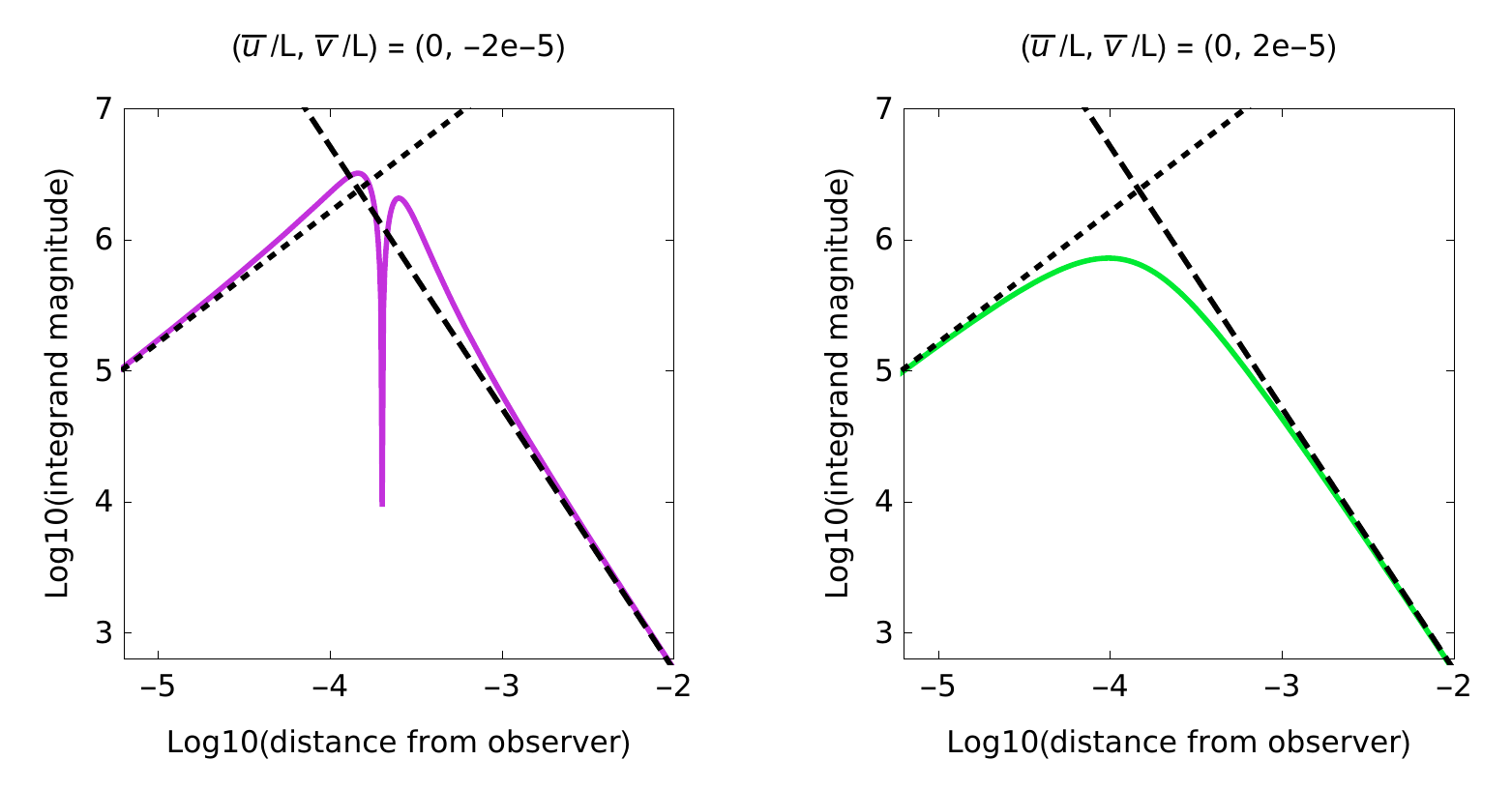}
    \caption{The $w$ component of the cusp acceleration integrands for two observers with $\bu=0$, one located at $\bv/L=-2\cdot10^{-5}$ in the past of the cusp (left panel, purple) and the other located at the same distance in the future of the cusp (right panel, green). In both plots, the short-dashed line indicates the predicted acceleration integrand when $u\ll\bu$, while the long-dashed line is for when $u\gg\bu$. Note that the left plot changes sign between regimes, while the right plot maintains the same sign throughout.}\label{fig:cusp-accelerations}
\end{figure}

Plots of the acceleration integrands for two observers near a cusp, demonstrating the phenomena discussed in this section, may be found in Fig.~\ref{fig:cusp-accelerations}. In order to obtain the solid lines from these plots, we carried out the calculation of the $w$-direction acceleration via Eqs.~(\ref{eqn:xuv},\ref{eqn:habg-num-den}) for $A$ and $B$ Taylor-expanded about an observer near a cusp on the Kibble-Turok loop~\cite{Kibble:1982cb}, keeping all terms up to fourth order in the lightcone constraint.

\subsection{Changes to the string near a cusp}\label{ssec:cusp-changes}

We have concluded that the acceleration as we approach a cusp diverges
like the inverse distance from the cusp to the observer (for the cusp
direction) or like the logarithm of the same (for the transverse
directions) and are only cut off by the near regime when $u$ is
comparable to this distance. On the other hand, the cusp is a
transient event which occurs at some precise $u$ and $v$ coordinates
on the worldsheet.  To find the total effect of back-reaction on a
point near a cusp due to the combined contributions of the rest of the
world sheet one should compute the change on the tangent vectors
following Eq.~(\ref{eqn:delta-nv}).  Upon integrating either of these
expressions, we will find that the $w$ direction is still divergent,
but only logarithmically, while the remaining directions are
non-divergent, and so both $\Delta A'$ and $\Delta B'$ will be
log-divergent in the $w$ direction.  Since the divergences are seen in
$\Delta A'$ and $\Delta B'$, they are not gauge artifacts.  As in the
kink case, integrating once again to determine the total loss of
length will give a finite answer.

The corrections $\Delta A'$ and $\Delta B'$ for a single oscillation
will never be large. For points very near the cusp, both corrections
will be proportional to $G\mu$ times a logarithm.  No logarithm
appearing in cosmology is more than about 100, and $100 G\mu$ is still
tiny for any realistic $G\mu$.

The only divergent correction is in the $w$ direction, which is the
direction of the cusp's motion, i.e., $A'_0 = B'_0$.  Nearby points
will have similar $A'$ and $B'$, so the correction acts mostly to
decrease the energy of the string near the cusp without changing the
directions of the tangent vectors.  Reparameterization to return
$\mathbf{A}'$ and $\mathbf{B}'$ to unit length will increase $A''$ and
$B''$, because $A'$ and $B'$ change by the same amount over less
parameter distance.  This decreases the strength of the cusp by
decreasing the area of the worldsheet in which $A'$ and $B'$ are
nearly identical.  The unit sphere looks more or less the same, but
the $A'$ and $B'$ now move more quickly over the cusp point, resulting
in weaker bursts of gravitational radiation in subsequent oscillations.

\section{Conclusions}\label{sec:conclusions}

We have demonstrated that points on a string worldsheet near a kink or
a cusp will feel a divergent acceleration due to those features. While
points not located at the feature itself always have some small nearby
region which looks smooth, divergent effects arise on a scale related
to the distance from that point to the nearby feature.

That there is a divergent acceleration as an observer approaches a
kink indicates that it is possible for the kink to be rounded off by
gravitational back-reaction, in contrast to the claim of
Ref.~\cite{Wachter:2016hgi} that kinks are ``opened'', and may seem
more similar to the ``smoothing'' of kinks used in
Ref.~\cite{Blanco-Pillado:2015ana}.  However, this rounding happens on
small distances at early times, and it takes a significant fraction of
the loop lifetime until a large length of string has been bent across
the kink. So while kinks are removed rapidly, the amount of string
spread across the gaps on the unit sphere is small.  Thus, cusps which
form as a consequence of this will be very weak.

Our results on back-reaction at cusps suggest that they lose a
significant amount of energy in the neighborhood of the cusp, making
them weaker as time passes. The effect of back-reaction will also
change the parameters that characterize the cusps, which could have
important consequences for their observational signatures.

These results were found using the zero-thickness string
approximation. Thus, once the observer approaches a kink or a cusp to
a scale comparable to the string thickness $\delta$, we expect the
expressions for the accelerations to change.\footnote{At that scale one
  would imagine that field theory effects of the type observed in
  simulations~\cite{Olum:1998ag} would be the dominant contribution to
  back-reaction.}  On the other hand, strings of cosmological and
astrophysical significance always have length scales many orders of
magnitude above their thicknesses,\footnote{For example: a Milky
  Way-scale string with $G\mu=10^{-11}$ has $L/\delta\sim10^{45}$.}
so these results are applicable to all but an infinitesimal fraction
of the string.

More importantly, the type of analysis done here is applicable only to
isolated, simple features on strings, and we can accurately calculate
only the initial effect.  After a significant period of back-reaction,
a string will have cusps that are partly depleted and look somewhat
like kinks, and kinks that are partly rounded and lead to weak cusps.
To fully understand the evolution of loops under the influence of
gravitational back-reaction, we need to numerically simulate
back-reaction over the course of the loop lifetime. We will report
on such simulations in future publications.

\section{Acknowledgments}

Concurrently to the work described here, Chernoff, Flanagan, and
Wardell~\cite{Chernoff:2018evo} did related work on cosmic string
back-reaction; that paper and this were submitted at the same time.
As far as we know the results are in agreement where they overlap.

We thank David Chernoff, \'Eanna Flanagan, Larry Ford, Mark Hertzberg,
Alex Vilenkin, and Barry Wardell for useful conversations.

This work was supported in part by the National Science Foundation
under grant numbers 1518742, and 1520792, the Spanish Ministry MINECO
grant (FPA2015-64041-C2-1P), and Basque Government grant (IT-979-16).
J. J. B.-P.  is also supported in part by the Basque Foundation for
Science (IKERBASQUE).

Finally, J. J. B.-P. and J. M. W. would like to thank the Tufts Institute
of Cosmology for its kind hospitality during the time that this work
was completed.

\bibliography{paper}

\end{document}